# Microfluidic front dynamic for the characterization of pumps for long-term autonomous microsystems

Yara Alvarez-Braña[1,2,◊], Andreu Benavent-Claro[3,4,◊], Fernando Benito-Lopez[2], Aurora Hernandez-Machado[3,4,*], Lourdes Basabe-Desmonts[1,5,*]

[1]Microfluidics Cluster UPV/EHU, BIOMICs Microfluidics Group, University of the Basque Country UPV/EHU, Vitoria-Gasteiz, Spain.

[2]Microfluidics Cluster UPV/EHU, Analytical Microsystems & Materials for Lab-on-a-Chip (AMMa-LOAC) Group, University of the Basque Country UPV/EHU, Vitoria-Gasteiz, Spain.

[3]Condensed Matter Physics Department, Physics Faculty, University of Barcelona, Spain

[4]Institute of Nanoscience and Nanotecnology (IN2UB), University of Barcelona, Spain

[5]Basque Foundation of Science, IKERBASQUE, Spain

◊ First co-authors

* Corresponding co-authors

**Highlights**

- Autonomous systems that operate for long periods of time, such as more than 20 hours.
- Mathematical model predicting the flow rate and operating times of the pumps.
- Universal architecture for flow control in autonomous microfluidic systems

**ABSTRACT**

To facilitate the use and portability of Lab on a chip technology, it is desirable to avoid the use of bulky electronic systems for flow control. Developed self-powered microsystems typically move only small volumes of fluid performing up to one or two hours. We have previously shown that polymeric micropumps combined with plastic microfluidic cartridges constitute a universal self-powered modular microfluidic architecture suitable for moving large volumes of fluids in short times. Herein, we show that polymeric micropumps can provide self-powered flow control for long periods of time in the range of hours and days. The calibration curves of various types of micropumps were obtained including one that maintained the movement of the fluid for 23 hours, advancing the fluid front up to 1.8 meters through a channel with a section of 0.127 mm$^2$. We found that the actuation time of the pump was related to degassing time, the effective surface area and the air recovery rate of the pump. This is the first example of a long operating self-powered microsystem, which may have a great impact in those cases where controlled flow is needed for a long period of time and the use of electronic equipment is not desirable.

# 1. INTRODUCCION

During the past decades, microfluidic technology has become an excellent tool for the development and manufacture of devices with applications in diverse fields [1,2]. The development of self-powered platforms is an area on the rise, as these autonomous devices can generate flow inside of the microfluidic channels without any source of external power, facilitating the transport and the use of these analytical platforms in the point of need [3]. Self-powered devices focus on fluidic control at a small scale for the movement of the sample and reagents during short periods of time, such as devices using finger-actuated pumps and valves [4], paper [5] or effervescent micropumps [6]. However, many microfluidic applications like cell culture (both for unicellular organisms [7] and mammal cells in cancer research or drug-development [8]), require precise flow control during long periods of time and they could benefit from long-term actuating autonomous microsystems. In addition, there are chemical microfluidics applications such as microreactor processes [9], droplet-based techniques [10], nanoparticle synthesis [11] or flow-through surface functionalization [12,13], that focused on the development of long-term controlled experiments; however, long-term operating autonomous devices are still an unmet goal in the field of microsystems [14].

In 2004, Hosokawa *et al*. [15] reported a power-free pumping methodology based on the use of degas driven flow for PDMS devices. After its degasification in a vacuum chamber, the high gas solubility of the PDMS enables this material to move the sample through the channel when it is placed at atmospheric conditions. Since then, degas driven PDMS devices have been used in the microfluidic field [16,17], leading to new strategies, such as the use of PDMS modular micropumps [18]. Our group recently demonstrated that polymeric cartridges of PDMS can act as a micropumps capable to absorb air capable to pull liquids and tested which material was most efficient for this use using a flowmeter [19,20]. In this study, we use front dynamics analysis [21,22] of a liquid inside of a microfluidic device, method that has been reported as a tool to study the capillary or pressure driven filling of microchannels to characterize the properties of fluids, such as the viscosity [23,24].

In this work, we aimed to evaluate the long-term performance of polymeric micropumps by analysing the degas-driven motion of the liquid front inside a long microchannel. The pressure generated by PDMS micropumps and the position and velocity of the fluid front were evaluated (**Figure 1**). For this, different pump parameters, such as degassing time, surface area and recovery rate were studied. Moreover, we could predict the behaviour of the pumps connected into a specific cartridge. When the pump is no-coated and pressure losses are generated, the pressure-driven diffusion model proposed by Woo *et al.* [25] can be used. When the pump is coated with epoxy resin to avoid the pressure loses, new expressions are proposed which mathematical model is obtained in [26]. From the obtained results, we would be able to predict the actuation time of the pump and flow rate of the sample and reagents inside of a specific microfluidic cartridge over time, by knowing the degasification time, the ratio between effective surface area and the volume of the PDMS pump and the dimensions of the channel, among other parameters.

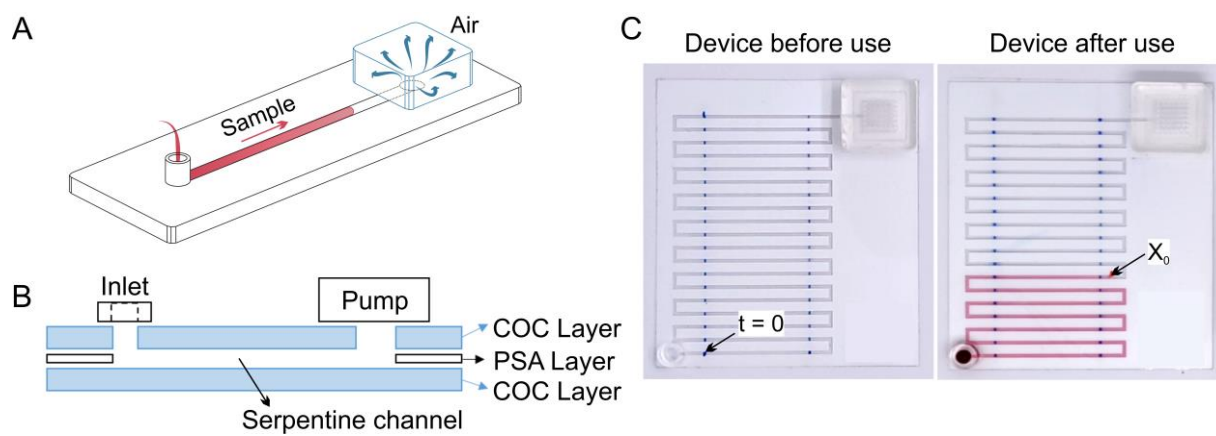

**Figure 1. Characterization of long-term PDMS micropumps.** A) Schematic representation of an operating autonomous microsystem using PDMS micropumps. B) Schematic side cut of the microfluidic system used to characterize PDMS micropumps over time, comprised of several assembled polymeric layers. C) Photographs of long serpentine cartridges employed to measure the advance of the front position of the liquid.

## 2. MATERIALS AND METHODS

### 2.1. PDMS micropumps fabrication and actuation

Different PDMS micropumps were fabricated by casting, using a previously described protocol [19]. For each pump a different mold was used to obtain the different effective surface areas (S = 683 mm$^2$, S1 = 290 mm$^2$ or S2 = 92 mm$^2$), **Figure SI1A**. In addition, to restrict the recovery of air through the external surfaces, pumps S, S1 and S2 were coated by a resin epoxy layer, only allowing the transfer of air through the effective surface area ($S_e$) in contact with the channel. The pumps were covered with a layer of liquid epoxy adhesive (Epoxy Araldite, Ceys) and placed in an *ad hoc* premade PMMA gasket adjusted to the pumps size (**Figure SI1B**) to prevent the exchange of air through the external surfaces. Therefore, the six manufactured pumps had different effective surface area ($S_e$), total surface ($S_t$) area and/or volume (**Table SI1**).

Manufactured PDMS micropumps were degassed in a RVR003H-01 vacuum chamber (Dekker Vacuum Technologies, USA) at 70 Pa during different degasification times (ranging from 5 to 180 min) and vacuum packed in a SV-204 vacuum sealer (Sammic, Spain) in order to store in airless environment until use. During the experiments, a rectangular-shaped PSA (146-µm thick ARcare® 90880, Adhesive Research, Ireland) was employed to assemble the micropumps to the microfluidic devices or the connector for the manometer measurements. These adhesives have been the subject of extensive study in the microfluidic field over the years [27] and have proven to be a useful material for microfluidic bonding in experiments achieving pressures of up to 1000 mbar [28].

**2.2. Fabrication of the multilayer-plastic devices for front fluidics characterization**

Polymeric microfluidic cartridges were fabricated by multiple-layer lamination. Microfluidic channels were cut by Graphtec cutting Plotter CE6000-40 (CPS Cutter Printer Systems, Spain) on sheets of Pressure Sensitive Adhesive layers (127-µm thick ARcare® 8939 white PSA, Adhesive Research, Ireland) and a 140-µm thick COC substrate (mcs foil 011, Topas, microfluidic ChipShop, Germany) was used as the bottom base and top layer of the device. All the cartridges included a 4-mm diameter inlet and a 10 mm x 5 mm (WxL) outlet at the end of the channel used as waste reservoir. The assembly and lamination of the COP and PSA layers, one on top of the other, generated the microfluidic serpentine of 0.127 mm height, 1 mm width and 900 mm length in total for the normal device; however, a longer channel 1800-mm long was used for the experiments using the epoxy-covered pumps. A 30-mm section was marked in the center of each serpentine segment as a point to measure the time when the front advanced through these sections. Finally, a round-shape PMMA piece (4 mm high and internal diameter) was connected to the inlet.

## 2.3. Front position and speed measurements

Vacuum-packed micropumps were opened and attached to the outlet of the microfluidic cartridge. After waiting 3 minutes, the samples (water-based solution of red food-dye) were loaded into the inlet of the device. The advance of the fluid front was recorded on video (OnePlus 6T rear cameras: 16 + 20 megapixels).

## 3. RESULTS AND DISCUSSION

### 3.1 Front flow dynamics on self-powered microfluidic system driven by PDMS non-coated micropumps (S, S1 and S2)

#### 3.1.1 Effect of the degasification time of the micropump in the fluid front dynamics

In order to investigate the effect of the degasification time of a modular PDMS micropump generating the flow inside of a channel, the front dynamics of a water-based fluid flowing through a self-powered microfluidic set up was optically monitored. The system consisted in a multilayer plastic cartridge (0,127 x 1 x 900 mm serpentine) connected to a PDMS micropump type S ($S_e$ = 683 mm$^2$) that was degassed during 1, 3, 5, 15, 30, 60, 120 or 180 minutes. During the first 30 minutes of experiment (1800 seconds), the front advanced rapidly with a maximum initial velocity between 0.31 mm s$^{-1}$ for a degasification time ($t_d$) of 1 min and 0.75 mm s$^{-1}$ for 180 min of degasification, and it gradually decreased until the fluid stopped in all cases. This initial velocity depended on the degassing time and, therefore, the actuation time of the pump ($t_a$) and the length of the channel traveled by the liquid until the front stopped ($X_0$) as well. The $t_a$ and the $X_0$ ranged from 9016 s (150 min) and 380 mm to 10706 s (178 min) and 895 mm for degasification times of 1 and 180 min, respectively.

Regarding the plot of the position of the front *versus* time, **Figure 2A**, we observed that the initial slope of the curve (the initial velocity of the fluid front) and the final position of the plateau ($X_0$) increased with the degasification time of the pump. The experimental data was fitted to equation (1) previously reported by Woo *et al*. [25].

$$X(t) = X_0 \left(1 - e^{-\frac{t}{\tau}}\right) \tag{1}$$

Where X is the front flow position through time, $X_0$ is the position where the front flow stopped and τ is a characteristic time related to the time it took to reach the plateau area of the curve. The calculated position of the front (lines) through the serpentine channel using a pump S fit correctly with the experimental data obtained (dots).

Sung Oh Woo *et al.* [25] showed that the $X_0$ value is a terminal distance that is determined by the channel dimensions and the ratio between effective surface area and the volume of the vacuum pocket inside the PDMS devices, however, when using our PDMS micropumps without vacuum-pocket this value also depends on the degasification time.

In addition, once the $X_0$ value was obtained, it was possible to extract the characteristic τ parameter from equation (1), **Table SI2**. Using several degasification times (15, 60, 120 and 180 min), it could be observed that the τ value for the pump S is around 20 min with a tendency to increase with the degasification time, that is, from 1298 seconds (21 min) for a 15 min-degassed micropump to 1581 seconds (26 min) for a 180 min-degassed micropump.

The longer the degasification time, the larger the volume of air is evacuated from the pump; and, therefore, the time for the pump to recover all the evacuated air would be longer as well. This increase in the activity time of the pump is consistent with the higher values of τ observed when the same micropump was degassed for longer times, thus, we can conclude that the activity time of the micropumps can be characterized by the parameter τ.

Using the definition of velocity as the derivative of position (equation 1) we obtain the following expression:

$$v(t) = \frac{dx}{dt} = v_0 \, e^{-\frac{t}{\tau}} \tag{2}$$

Where $v$ is the velocity of the fluid front and $v_0 = X_0/\tau$ is the initial velocity of the front. Finally, to obtain the expression for the flow rate of the fluid (equation 3), we used the definition of flow rate in a constant cross section channel and the definition of front velocity (equation 2):

$$Q(t) = A_{ch}\frac{dx}{dt} = Q_0\, e^{-\frac{t}{\tau}} \qquad (3)$$

Where $Q$ is the flow rate of the fluid front; $A_{ch}$, the cross section of the channel where the measurements are made; and $Q_0 = \frac{A_{ch}X_0}{\tau}$ is the initial flow of the front. In order to evaluate if this model agrees with the experimental results, front velocity and flow rate of the liquid were obtained for each serpentine section. **Figure 2B** shows the calculated flow of the front (lines) and the experimental values obtained for the front velocity and front flow in each serpentine section (dots) for 15 min of degasification time of the pump S. As it can be observed, for pumps subjected to identical degassing times, both the front velocity and fluid flow exhibited a consistent trend throughout the experiment, showcasing an exponential decay as anticipated and causing the fluid to move slower as it goes through each section of the serpentine. However, we observed that the overall front velocity and flow rate increased with the degasification time. That is, using a 15-min degassed micropump, the velocity of the front and the fluid flow rate were 0.25 mm s$^{-1}$ and 31 nL s$^{-1}$, respectively. In contrast, using the same 60-min degassed micropump resulted in a front velocity of 0.30 mm s$^{-1}$ and a flow rate of 38 nL s$^{-1}$ after 12 min of experiment.

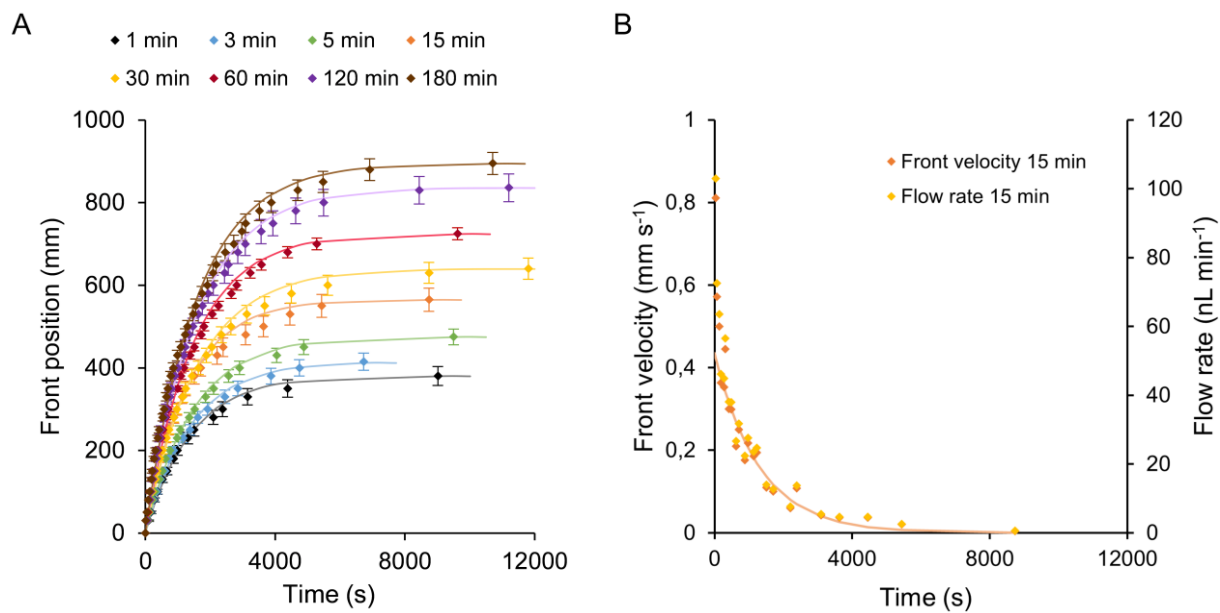

**Figure 2. Effect of the degasification time in the fluid front dynamics.** A) Plot of the front position of the sample *versus* time for pump S degassed during different degasification times between 1 to 180 minutes. B) Plot of the front velocity and the flow rate of the sample *versus* time produced by an unimodular micropump S degassed for 15 minutes. The dots show the experimental data, and the lines show the calculated data using the corresponding equation in each case.

### 3.1.2 Effect of surface area of the micropump on fluid front dynamics

The micropumps used in these experiments have 6 sides, where the effective surface area ($S_e$, the side exposed to the microchannel) is the main driver force inside of the microdevices. During this experiment, the front dynamics of a water-based fluid flowing through a serpentine microchannel using three pumps of similar volume and different $S_e$ was monitored.

The three different types of micropumps [S ($S_e = 683$ mm$^2$), S1 ($S_e = 290$ mm$^2$) and S2 ($S_e = 92$ mm$^2$)] were degassed during 60 or 180 min and, then, connected to the outlet of the plastic cartridge to trigger the movement of the fluid inside the microchannel (**Figure 3**). It was observed that, for a fixed degasification time ($t_d$), a decrease in the $S_e$ (S > S1 > S2) resulted in a smaller $X_0$ value and a smaller fluid front velocity, producing an increase in the actuation time of the pump ($t_a$) and, consequently, an increase in the value of $\tau$.

In view of the results obtained, when generating the same pumping pressure, decreasing $S_e$ reduces the amount of gas molecules that can enter the PDMS micropump at the same time. Therefore, as the front fluid dynamics is directly dependent on the amount of gas that the pump can absorb from the close microfluidic channel, given a fixed degasification time, the final front position of the liquid and the velocity of the front decreased with the $S_e$ of the pump.

In addition, the effect of the different $t_d$ and $S_e$ on the $X_0$ and $\tau$ value could be compared (**Table SI3**). On the one hand, a 200% increase in the $t_d$ of pump S (from 60 to 180 min) produced a 9% increase in $\tau$ value, while a 135% increase in the $S_e$ (from S1 to S) for the same degassing conditions (60 min) produced a 98% increase of $\tau$ value. Thus, variating the $S_e$ had a bigger impact on the $\tau$ value than changing the degasification conditions. On the other hand, a 200% increase in the $t_d$ of pump S (from 60 to 180 min) produced a 24% increase in $X_0$ value, while a 136% increase in the $S_e$ (from S1 to S) for the same degassing conditions (60 min) produced a 12% increase. In this case, the effect produced by the different $S_e$ and $t_d$ had a similar impact on the $X_0$ value.

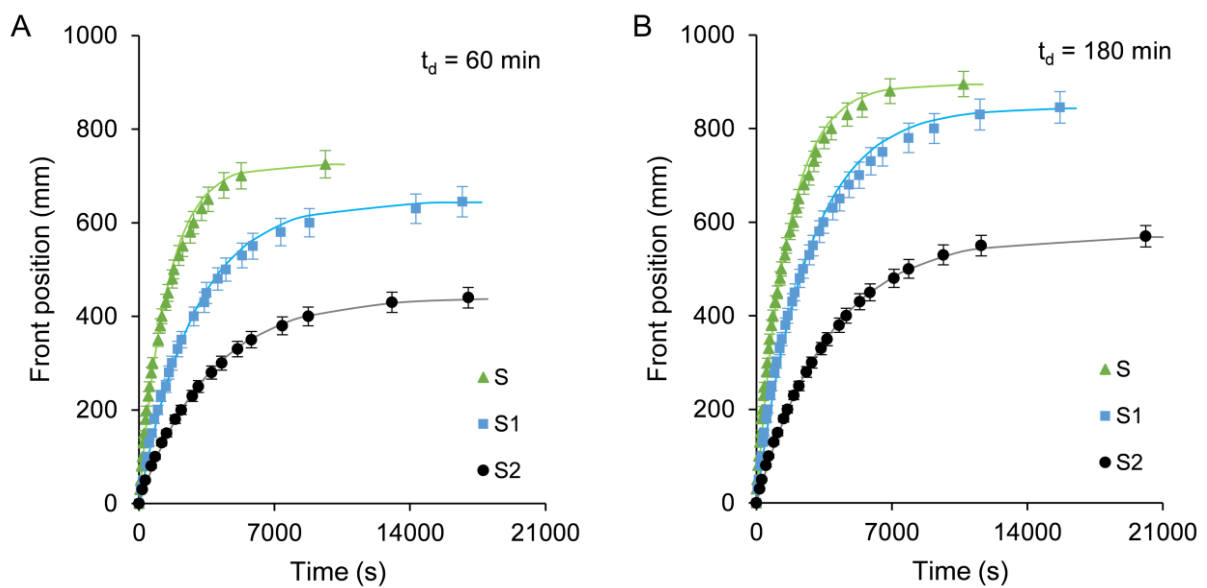

**Figure 3. Effect of surface area of the micropump on fluid front dynamics.** Plot of the fluid front position *versus* time using pumps S, S1 or S2 degassed during the same degasification time: A) 60 min or B) 180 min. The dots show the experimental data, and the lines show the calculated data using equation (1) in each case.

**3.2 Front flow dynamics on self-powered microfluidic system driven by PDMS epoxy-coated micropumps (EPX-S, EPX-S1 and EPX-S2)**

**3.1.1 Effect of the degasification time of the micropump in the fluid front dynamics**

After characterizing the non-coated PDMS pumps, in order to investigate the effect of the degasification time when an epoxy-coated PDMS micropump generated the flow inside of a channel, the front dynamics of a water-based fluid flowing through a self-powered microfluidic set up was optically monitored. In this case, the system consisted in a larger multilayer plastic cartridge (0,127 x 1 x 1800 mm serpentine) connected to an epoxy-coated PDMS micropump type S (EPX-S) that was degassed during 5, 15 or 30 minutes.

The epoxy-coated pumps exhibit a similar behavior to the non-coated pumps when connected to the microfluidic device. In both cases, the front of the liquid advances rapidly through the channel, and the velocity of the fluid inside the microchannel gradually decreases over time until it eventually stops. However, a notable difference is observed with the epoxy-coated pumps, as they do not experience leakage through their external surfaces. Therefore, the non-coated micropump was saturated with air more quickly than in the case of pumps with epoxy coated counterpart. The $t_a$ difference between the coated (**Figure 4**) and then non-coated pump S (**Figure 2**) was more than 19 hours, so the effect of the external surfaces should be considered. This result highlights the effectiveness of epoxy coating in improving the functionality and efficiency of microfluidic pumps, offering advantages such as prolonged activity and reduced risk of leakage in practical applications.

The experimental data for the non-coated pumps was fitted to equation (1), however, the data of the epoxy-coated pumps could only be fitted during the first minutes of the experiment (short-term times). Hence, for these pumps, the experimental results could be fitted using equation (4), which is mathematically derived in Ref. [26]:

$$X(t) = X_0 \left(1 - e^{-\frac{t}{\tau}}\right)^{\frac{1}{2}} \quad (4)$$

Where the square root is fitted using a power law. It has been proved experimentally that all pumps with an external epoxy-coated layer obey this expression. Then, the $\tau$ value corresponding to the different micropumps was calculated using equation (1) for the non-coated and equation (4) for the epoxy-coated ones (**Table SI4**). In this case, the difference between the $\tau$ value of the epoxy-coated and non-coated pumps was significant, however, this was related to the different fitting equations used in each case due to their different actuation times.

In addition, as in the non-coated case, to obtain the expression for the front flow (equation 6), we derive the front position model (equation 4) to obtain the front velocity (equation 5) and we multiply it for the cross section of the channel:

$$v(t) = \frac{dx}{dt} = v_0 \frac{e^{-\frac{t}{\tau}}}{\left(1 - e^{-\frac{t}{\tau}}\right)^{\frac{1}{2}}} \quad (5)$$

Where $v$ is the velocity of the fluid front and, in this case, $v_0 = X_0 / 2\tau$ is a characteristic velocity.

$$Q(t) = A_{ch} \frac{dx}{dt} = Q_0 \frac{e^{-\frac{t}{\tau}}}{\left(1 - e^{-\frac{t}{\tau}}\right)^{\frac{1}{2}}} \quad (6)$$

Where $v$ is the velocity of the fluid front and, in this case, $Q_0 = \frac{A_{ch} X_0}{2\tau}$ is a characteristic flow.

As an example, **Figure 4B** shows the expected flow rate using equation (6) (lines) and the experimental values obtained for the front velocity in each serpentine section (dots) for 30 min of degasification time.

Equations (4) and (6) provides a method for understanding and predicting the performance of epoxy-coated pumps under various operating conditions.

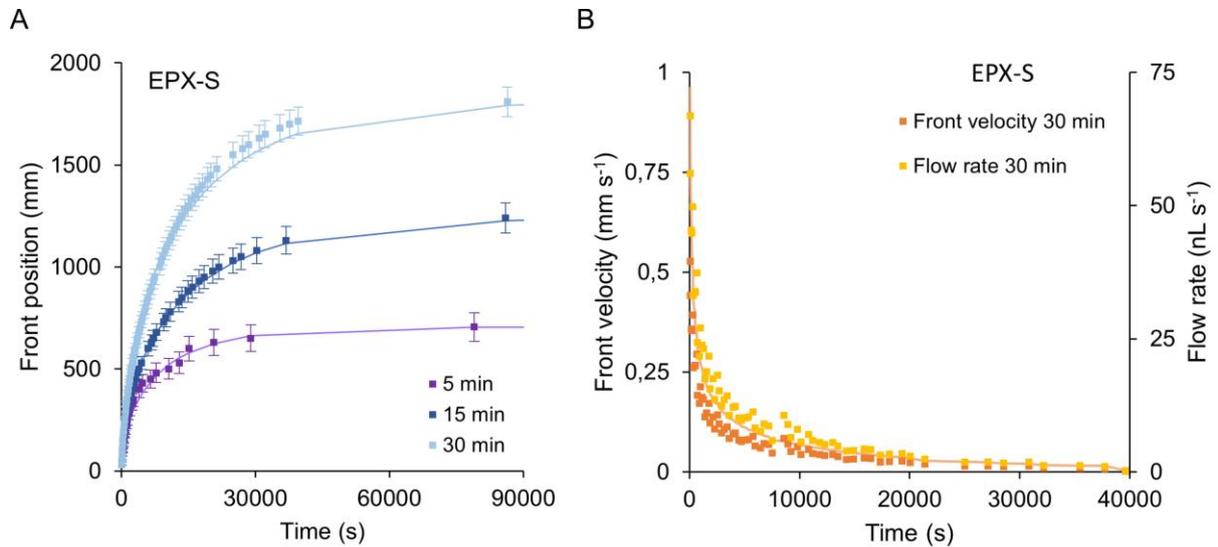

**Figure 4. Effect of the degasification time in the fluid front dynamics for an epoxy-coated pump.** A) Plot of the front position of the sample *versus* time for pump EPX-S degassed during 5, 15 and 30 minutes. B) Plot of the front velocity of the sample *versus* time produced by an unimodular epoxy-coated micropump degassed for 30 minutes. The dots show the experimental data, and the lines show the calculated data using the corresponding equation (6).

### 3.1.2 Effect of surface area of the micropump on fluid front dynamics

During this experiment, the front dynamics of a water-based fluid flowing through a serpentine microchannel using three epoxy-coated pumps of similar volume and different $S_e$ was monitored.

The three different types of micropumps [EPX-S ($S_e$ = 683 mm$^2$), EPX-S1 ($S_e$ = 290 mm$^2$) and EPX-S2 ($S_e$ = 92 mm$^2$)] were degassed during 15 or 30 min and, then, connected to the outlet of the plastic cartridge to trigger the movement of the fluid inside the microchannel (**Figure 5**).

It was observed a similar behavior than in the case of the non-coated micropumps, that is, for a fixed degasification time ($t_d$), a decrease in the $S_e$ resulted in a smaller $X_0$ value and a smaller fluid front velocity, producing an increase in the actuation time of the pump ($t_a$) as we mention in the previous section.

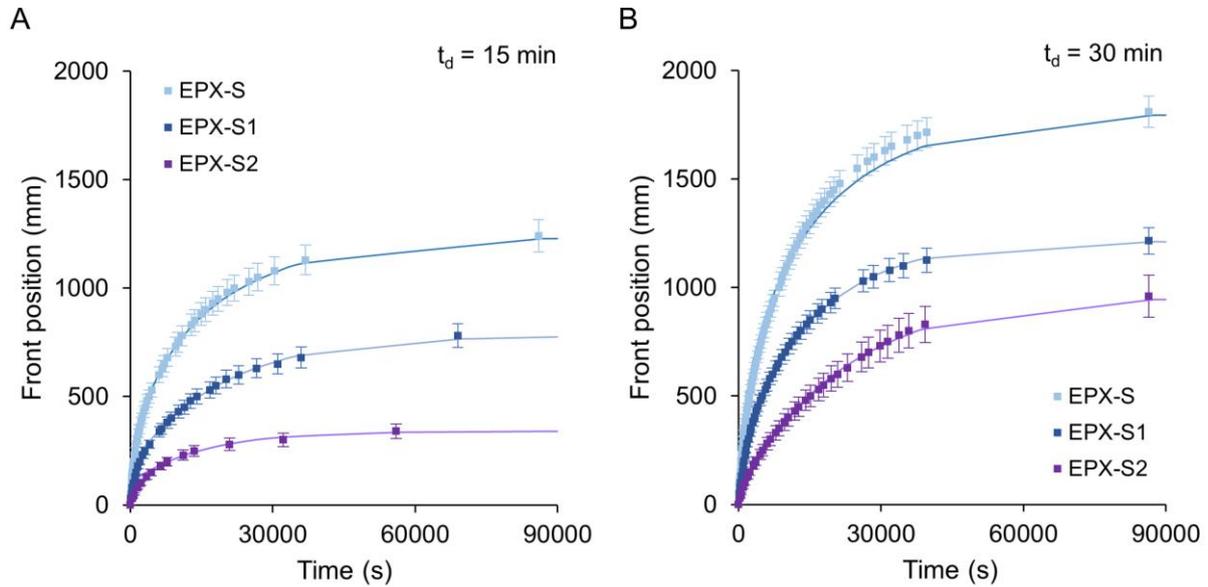

**Figure 5. Effect of surface area of the micropump on fluid front dynamics for the epoxy-coated micropumps.** Plot of the fluid front position versus time using pumps EPX-S, EPX-S1 or EPX-S2 degassed during the same degasification time: A) 15 min or B) 30 min. The dots show the experimental data, and the lines show the calculated data using equation (3) in each case.

**3.3 Effect of the air recovery rate of the micropump on fluid front dynamics: Comparison of non-coated and epoxy-coated micropumps (S, S1 and S2 *vs*. EPX-S, EPX-S1 and EPX-S2)**

To evaluate the time that the pump takes to reach the equilibrium again after vacuum (recovery rate), non-coated micropumps (S, S1 and S2) and epoxy-coated micropumps (EPX-S, EPX-S1 and EPX-S2) were degassed for 15 or 30 min and connected to the outlet of a 1800 mm-long serpentine channel.

In the experiment it was observed that, non-coated pumps reached the equilibrium up to a plateau more quickly than the epoxy-coated pumps, which achieved a larger $X_0$ value and actuation time ($t_a$) than their non-coated counterparts (for instance, the epoxy coating of pump S generated a 180% increase in $X_0$ value when degassed during 30 min). This indicates that, in the case of the non-coated PDMS pumps, the air absorbed by the external surfaces affects its air recovery rate.

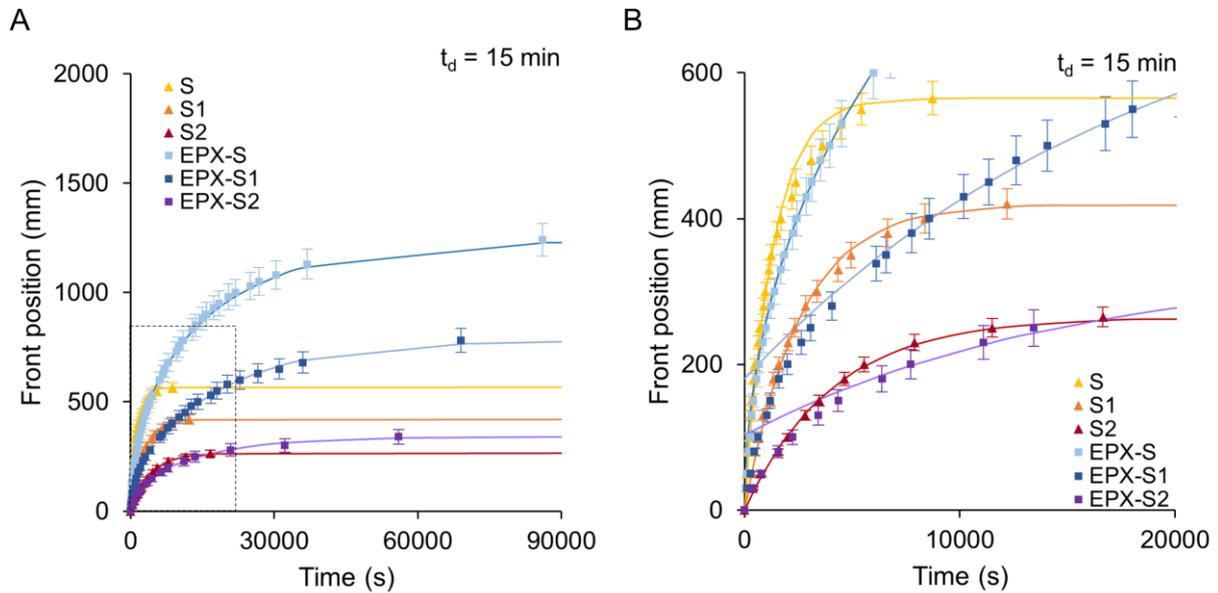

**Figure 6. Effect of the recovery rate of the micropump on fluid front dynamics.** Plot (A) and magnification of the first 30 min (B) of the front position of the fluid *versus* time, driven by pumps S, S1, S2, EPX-S, EPX-S1 and EPX-S2 degassed during 15 min. The dots show the experimental data, and the lines show the calculated data using the corresponding equation in each case.

**CONCLUSIONS**

PDMS degassed micropumps enable the autonomous flow of samples into microfluidic systems. Understanding the behavior of the system will allow the custom design of micropumps for personalized long-term activity microfluidics. The experimental data showed a similar tendency for the advance of the fluid front and the manometer pressure for all proposed micropumps, demonstrating that, in all cases, the position of the front increases until reaching a plateau as the pump enters a stationary state.

We determined that the actuation times and the total amount of fluid moved by the micropumps are related to its effective surface area, the degasification time and the recovery rate. Therefore, the customizability properties of these micropumps allows them to be used in different scenarios with different pressure and activity time requirements. In addition, by characterizing the pumps and measuring the $X_0$ and the $\tau$ value through the analysis of the front dynamics of a fluid inside a microfluidic channel, we could predict the flow rate and the actuation time of a pump connected to a device, generating a controlled and customized self-powered microsystem for short or long-term applications.

Lastly, unlike current systems, these PDMS micropumps can generate flow for extended periods (up to 23 hours using epoxy-coated pumps), facilitating the design of autonomous microfluidic devices for the movement of large volumes of sample with an estimation and control of the flow during the whole experiment. For example, these architectures could sustain cell cultures by supplying nutrients without electronic equipment and minimal user intervention during days.

ACKNOWLEDGMENTS

YA-B, FB-L and LB-D acknowledge funding support from "Ministerio de Ciencia y Educación" (Spain) under grant PID2020-120313GB-I00/AIE/10.13039/501100011033 and from Basque Government under ''Grupos Consolidados'' with Grant No. IT1633-22.

AB-C and AH-M acknowledge support from "Ministerio de Ciencia, Innovación y Universidades" (Spain) under project PID2022-137994NB-100 and AGAUR (Generalitat de Catalunya) under project 2021-SGR-00450.

We all acknowledge the support of the Spanish Ministry of Economy and Competitiveness through grant DPI2015-71901-REDT (MIFLUNET), partly funded through European Funds.

# Supporting information: Microfluidic front dynamics for the characterization of pumps for long-term autonomous microsystems.

Yara Alvarez-Braña[1,2,◊], Andreu Benavent-Claro[3,4,◊], Fernando Benito-Lopez[2], Aurora Hernandez-Machado[3,4,*], Lourdes Basabe-Desmonts[1,5,*]

[1]Microfluidics Cluster UPV/EHU, BIOMICs Microfluidics Group, University of the Basque Country UPV/EHU, Vitoria-Gasteiz, Spain.

[2]Microfluidics Cluster UPV/EHU, Analytical Microsystems & Materials for Lab-on-a-Chip (AMMa-LOAC) Group, University of the Basque Country UPV/EHU, Vitoria-Gasteiz, Spain.

[3]Condensed matter physics department, Physics Faculty, University of Barcelona, Spain

[4]Institute of Nanoscience and Nanotecnology (IN2UB), University of Barcelona, Sapin

[5]Basque Foundation of Science, IKERBASQUE, Spain

◊ First co-authors

* Corresponding co-authors

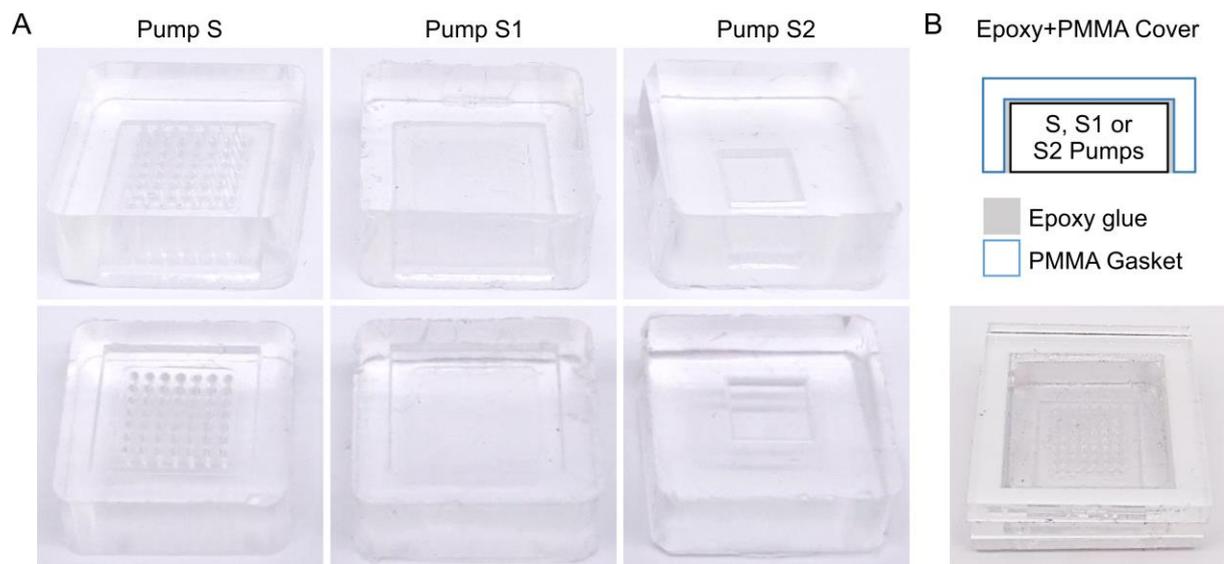

**Figure SI1. PDMS micropumps.** A) External surface (top) and effective surface area (bottom) photographs of micropumps S, S1 and S2. B) Scheme of the layers and top photograph of the epoxy + PMMA cover used to fabricate the micropumps EPX-S, EPX-S1 and EPX-S2 by covering the external surface area of micropumps S, S1 and S2.

**Table SI1.** Specifications of the six PDMS micropumps.

| Type of Pump | PDMS bulk volume (V) (µL) | Effective surface ($S_e$) (mm²) | $S_e$/V Ratio | Total surface ($S_t$) (mm²) | $S_t$/V Ratio |
|---|---|---|---|---|---|
| S | 4052 | 683 | 0.1686 | 1959 | 0.4835 |
| S1 | 4141 | 290 | 0.0700 | 1566 | 0.3782 |
| S2 | 4298 | 92 | 0.0214 | 1368 | 0.3183 |
| EPX-S | 4052 | 683 | 0.1686 | 683 | 0.1686 |
| EPX-S1 | 4141 | 290 | 0.0700 | 290 | 0.0700 |
| EPX-S2 | 4298 | 92 | 0.0214 | 92 | 0.0214 |

**Table SI2.** $X_0$ and τ value calculated for 15, 60, 120 and 180 min of degasification time of pump S.

| Degas. Time (min) | $X_0$ value (mm) | τ value (s) |
|---|---|---|
| 15 | 565 | 1298 |
| 60 | 725 | 1471 |
| 120 | 836 | 1576 |
| 180 | 895 | 1581 |

**Table SI3.** Experimental $X_0$, experimental $t_a$ and calculated $\tau$ value for pumps S, S1 and S2 after 60 or 180 min of degasification.

| Pump | Effective surface $(S_e)$ (mm$^2$) | Degas. time (min) | $X_0$ value (mm) | $\tau$ value (s) | $t_a$ (min) |
|------|------|------|------|------|------|
| S  | 683 | 60  | 725 | 1439 | 160 |
| S1 | 290 | 60  | 645 | 2854 | 278 |
| S2 | 92  | 60  | 440 | 3605 | 284 |
| S  | 683 | 180 | 895 | 1574 | 178 |
| S1 | 290 | 180 | 845 | 2667 | 261 |
| S2 | 92  | 180 | 570 | 3781 | 335 |

**Table SI4.** Experimental $X_0$ and calculated $\tau$ for pumps S, S1, S2, EPX-S, EPX-S1 and EPX-S2 after 15 or 30 min of degasification.

| Pump | Degas. time (min) | $X_0$ value (mm) | $\tau$ value (s) |
|------|------|------|------|
| S      | 15 | 565  | 1298  |
| S1     | 15 | 420  | 2585  |
| S2     | 15 | 265  | 4067  |
| EPX-S  | 15 | 1240 | 22172 |
| EPX-S1 | 15 | 780  | 27280 |
| EPX-S2 | 15 | 340  | 20424 |
| S      | 30 | 640  | 1594  |
| S1     | 30 | 540  | 2753  |
| S2     | 30 | 340  | 3658  |
| EPX-S  | 30 | 1810 | 22770 |
| EPX-S1 | 30 | 1215 | 24695 |
| EPX-S2 | 30 | 960  | 35000 |